\documentclass[12pt]{article}   	
\usepackage{geometry}                		
\usepackage[braket, qm]{qcircuit}
\usepackage{graphicx}				
\usepackage{amssymb, setspace, amsmath, pgfplots}
\usepgfplotslibrary{groupplots}
\usetikzlibrary{pgfplots.groupplots}
\usetikzlibrary{plotmarks}

\pgfplotsset{compat=1.5, legend style={font=\footnotesize}}
\usepackage[size=small,labelfont=bf]{caption}
\usepackage{algorithm, algorithmic}
\usepackage{url}
\usepackage{authblk}
\date{}							
\title{\vspace{-10mm} Recursive Path-Summing Simulation of \\ Quantum Computation\vspace{-2mm}}
\author{Andrew Shi \vspace{-2mm} \\ Henry M. Gunn High School \\ 
780 Arastradero Road, Palo Alto, CA 94306, USA \\
\texttt{{theandrewshi@gmail.com}}}
\begin{document}
\maketitle
\vspace{-10mm}
\begin{onehalfspacing}
\begin{abstract}
Classical simulation of quantum computation has often been viewed as the method to determine where the horizon of quantum supremacy is located---that is, where quantum computation can no longer be simulated by classical methods. As of now, the 50 qubit threshold for quantum supremacy has been determined largely by the state vector simulation method's exponential space demands placing an upper bound on simulation memory capabilities. To investigate this claim, we present and test an implementation of a known path integral simulation algorithm running in linear space; the method is based on recursively traversing the underlying computation tree for quantum algorithms and summing over possible amplitudes. We find that the implementation is able to simulate the hidden subgroup method (HSP) standard method---a notable class of circuits including Shor's algorithm amongst others---in a reasonable amount of time using extremely low memory, as well as other circuits with similar parameters. The performance results of this algorithm suggest that it can serve as a feasible alternative to state vector simulation and that with respect to the HSP standard method, quantum supremacy may be more accurately measured using the recursive path-summing method on large numbers of qubits, compared to the state vector method.
\end{abstract}
\end{onehalfspacing}

\section{Introduction}
	In the past few decades, interest in quantum computation has begun to steadily grow as a result of ideas such as Richard Feynman's concept of simulating physics with a computer operating on the laws of quantum physics \cite{feynman}. Since then, the field has been explored by many quantum theorists, yielding unusually exciting results such as Shor's polynomial-time factoring algorithm \cite{shor} demonstrating the capability of quantum computation to provide dramatically improved solutions to important mathematical and computational problems. Recently, quantum computing has transitioned from the theoretical drawing board to the laboratory as technological advances have enabled construction of the first quantum computers \cite{smallquantcomp}; with the production of experimental devices, there is a real possibility of quantum computing becoming the new paradigm of computational power. Therefore, the notion of ``quantum supremacy'' was developed to reference the possibility that after a certain point, quantum computation may be more powerful than all current classical computation. To this end, classical simulation is viewed as an accurate measurement of where this ``supremacy horizon'' may lie by considering the greatest number of qubits classical machines are able to simulate \cite{supremacy}.
\par
	Due to quantum algorithms' peculiar properties, simulation algorithms usually run in exponential space and time with respect to the number of qubits involved, exorbitant resource requirements which limit simulations to a size of around $\sim$ 40 to 50 qubits on present day state-of-the-art supercomputers \cite{largesim}. This limitation can be seen in the most prevalent and widely used generic simulation methods: the state vector simulation approach. In quantum computers, a single unit of information is stored in a qubit, the quantum analogue of a classical bit which takes the form $\alpha|0\rangle + \beta|1\rangle$ for complex numbers $\alpha, \beta$ satisfying $|\alpha|^2 + |\beta|^2 = 1$; because each qubit has a probability to be in either the $|0\rangle$ or $|1\rangle$ states, a system of $n$ qubits has an associated amplitude for each of the $2^n$ possibilities for the classical $n$-bit register. Thus, the state vector algorithm records the entire $2^n$-size state space of a $n$-qubit system in memory as a vector of state amplitudes and updates the vector according to the application of each gate in the quantum circuit model (i.e.\ a representation of a quantum algorithm). Applying each gate to the state vector requires iteration over all $2^n$ elements of the vector: therefore, for a circuit of $l$ gates, the time and space complexities of this algorithm are $O(l2^n)$ and $O(2^n)$, respectively. Therefore, as mentioned before, this simulation method surpasses 1-terabyte memory usage for simulations on $>$40 qubits, rendering simulations extremely costly; there have been few subsequent endeavors to improve the simulation of general quantum circuits in either time or space, and many have accepted---based on the state vector method's memory requirements---that current classical computers are only able to simulate a maximum of 50 qubits, defining a concrete marker for quantum supremacy.
\par
	The state-vector algorithm's lack of optimizable structure along with previously discovered quantum algorithms yielding exponential resource savings over their classical counterparts has led many quantum scientists to assume that quantum computation is inherently more powerful than classical computation, thus requiring exponential resources at minimum to classically simulate. To further examine the details of this supremacy assumption with respect to time, another direction in simulation research focuses on finding specific subclasses of quantum circuits simulable in sub-exponential time, exploiting structure arising from a common feature of the subclass to reduce resource usage. As a result, many interesting subcategories of circuits have been discovered to use sub-exponential time for classical simulation \cite{stabilizer, limitedentangle, quidd, tensor, pfaffian}; in particular, the Gottesman-Knill stabilizer representation \cite{stabilizer} expanded upon by Aaronson and Gottesman \cite{newstabilizer} uses $O(n^3)$ time to simulate a large group of circuits based on the specific, efficiently-simulatable properties of stabilizer states. However, while similar non-general simulation algorithms can be interesting in their own right by investigating the source of quantum computing's power and testing specific cases, many interesting quantum circuits do not fall within these algorithms' narrowly defined groups, in which case quantum scientists must resort to all-purpose simulation. Furthermore, they focus on reducing time requirements for simulation, which has little effect on the memory-restricted supremacy horizon.
\par
	Seeing that the quantum supremacy boundary of 50 qubits is based primarily on the memory restrictions of the state vector algorithm, another important question is raised: is it possible to implement a general purpose simulation method with sub-exponential memory complexity, which could determine a more accurate quantum supremacy boundary? In this paper, an simulation implementation adapted from a commonly known algorithm (the path integral method) based on traversing paths sequentially to minimize space usage is tested to show that this may be the case for certain circuits, some of which have demonstrated the most exaggerated quantum vs. classical time complexity speedups known to date. We will show that the experimental results notably apply to Shor's algorithm \cite{shor}, the polynomial-time factoring algorithm with major future applications and value. 
\par
	The rest of the paper features various analyses of the algorithm and empirical studies of its performance demonstrating resource requirements. Section 2 describes the recursive path-summing simulation algorithm running in exponential time and linear space, space-time complexity analysis for the algorithm, and the only C++ implementation of the algorithm known to date (the PocketSimulator project). Section 3 analyzes several example executions' time and space usages in addition to comparing space-time complexities between PocketSimulator, another simulation algorithm with similarly small space requirements, and the state vector approach using multiple interesting circuits. Finally, the implications of this algorithm for quantum computing and quantum supremacy are mentioned in section 4, along with future directions to move in and unanswered questions.
	
\section{The Algorithm}
	One angle of approaching quantum computation is to view the action of a quantum circuit on an input as a mapping each of the $2^n$ classical states and their corresponding amplitudes to one \textit{or more} classical states, with the case of \textit{multiple} mapped states made possible by the quantum phenomenon of superposition. According to this approach, quantum gates may be classified as ``branching'' and ``non-branching'' based on whether or not they are capable of mapping one classical state to multiple (i.e.\ producing a superposition of states); this interpretation of quantum computation will be used to explain the algorithm, which simulates a quantum circuit by calculating the amplitude \melem{endState}{C}{startState} for a user-chosen circuit $C$ and values of $endState$/$startState$ (the desired ending/starting states of the qubit register).
\par
	A quantum gate which acts non-trivially on $n$ qubits can be represented by a square $2^n$ by $2^n$ unitary matrix; if the gate matrix has only one entry per row, then the gate is defined as ``non-branching''---that is, when operating on any $n$-bit classical state, it is only capable of flipping specific bits and/or adding a relative phase factor to the state. ``Non-branching'' gates include the identity, Pauli gates, CNOT, and Toffoli gates, because each of these gates map a given classical state to another single classical state (while possibly introducing a phase factor that adjusts the state's amplitude). However, when the gate matrix has more than one entry per row, then the gate is ``branching''. Such gates are capable of mapping a given state to a superposition of quantum states; a simple and common example is the Hadamard quantum gate: 
	
\begin{onehalfspacing} $$ H =  \frac{1}{\sqrt{2}} \begin{bmatrix} 1 & 1 \\ 1 & -1 \end{bmatrix} $$ 
$$ H|0\rangle = \frac{1}{\sqrt{2}} \begin{bmatrix} 1 & 1 \\ 1 & -1 \end{bmatrix} \begin{bmatrix} 1 \\ 0 \end{bmatrix} = \frac{1}{\sqrt{2}} \begin{bmatrix} 1 \\ 1 \end{bmatrix} = \frac{1}{\sqrt{2}}(|0\rangle + |1\rangle) $$ 
$$ H|1\rangle = \frac{1}{\sqrt{2}} \begin{bmatrix} 1 & 1 \\ 1 & -1 \end{bmatrix} \begin{bmatrix} 0 \\ 1 \end{bmatrix} = \frac{1}{\sqrt{2}} \begin{bmatrix} 1 \\ -1 \end{bmatrix} = \frac{1}{\sqrt{2}}(|0\rangle - |1\rangle) $$ 
\end{onehalfspacing} 

\noindent As shown above, the Hadamard gate takes each of the $|0\rangle$ and $|1\rangle$ basis states into an equal superposition of both basis states, and therefore ``branches'' a given $n$-bit state into two others. In the function mapping interpretation, ``branching gates'' map a state onto two or more distinct states which each have an amplitude calculated based on the previous gate's matrix representation. 
\par	
	Based on this interpretation of quantum gates, an implementation of the path integral simulation method can be derived based on the circuit graph representation. The first gate of the circuit is drawn as a single node, labeled with the initial state of the $n$-qubit register---presumably a single classical state. If this first gate is a ``non-branching'' gate, a single edge is drawn representing the single mapping of the gate on the initial state. If the gate is ``branching'', then two or more edges are drawn corresponding to the number of states the initial state was mapped to as a result of the gate's action: for example, for the application of a Hadamard gate on the first qubit, two edges would be drawn: one corresponding to the possibility where the first qubit becomes a 0, and the other corresponding to when the first qubit becomes a 1. Next, at the end of each edge, another node is drawn and labeled with the single classical state produced as a result of traveling down that edge. All nodes drawn at the end of this step represent the states produced by the first gate's mapping from the initial state; the rest of the tree is then drawn by applying the same procedure for each node, producing a representation of the quantum circuit's action. An example circuit on 3 qubits and its corresponding tree are shown in Figure 1.
\begin{figure}[bt]
\centering
\includegraphics[width=0.6\textwidth]{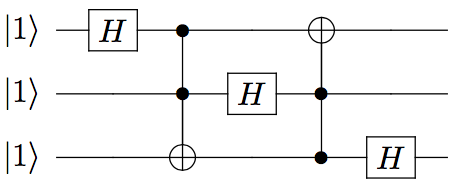}
\includegraphics[width=0.55\textwidth]{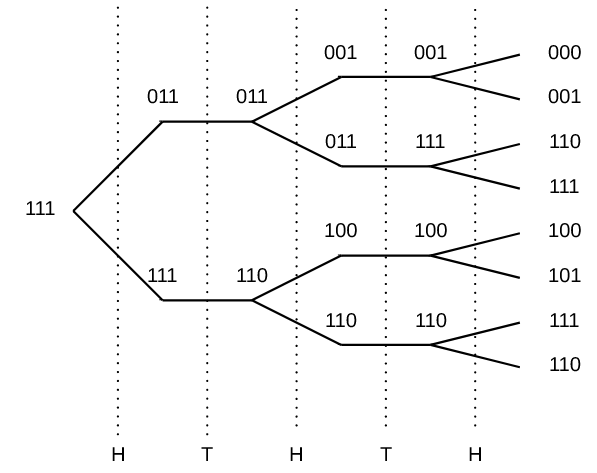}
\caption{a simple quantum circuit on 3 qubits and its corresponding tree representation. The tree diagram flows chronologically from left to right and shows the application of each gate. Different phase factors on each edge are not shown.}
\end{figure}

\par
	From such a graph representation of a quantum circuit, the recursive path-summing algorithm can be illustrated intuitively. Beginning from the single node on the left side of the tree diagram, the algorithm computes and records the accumulated phase and state changes of any ``non-branching'' gates until a ``branching'' gate is reached at a node. Then, it recursively calls itself for each edge leading out from that node, calculating the amplitude of reaching $endState$ from each of the branched nodes one at a time and adding them together (multiplied by respective transition phase factors specified in the ``branching'' gate matrix). Finally, an instance of the algorithm which reaches the end of the tree returns an amplitude of 1 if the instance's tracked state equals the desired $endState$ and 0 otherwise, in order to include only paths which reach $endState$ in the overall calculation of \melem{endState}{C}{startState}.
\par	
	Note that this procedure always completely calculates the amplitude of one edge before moving on to the next, effectively ``traveling'' down a possible branch and returning to explore others later, in the same order as a depth-first search. In this fashion, the intermediate amplitudes of recursive calls trickle backwards along the tree, accumulating phase factors from circuit gates and added to amplitudes of other branches until the the algorithm calculates a single value for $\langle endState|C|startState \rangle$, obtained by summing over all the recursively enumerated tree paths.
\par 
	Now we prove the space complexity of $O(n+h)$ for simulating a circuit of $t$ non-branching gates and $h$ branching gates on $n$ qubits by examining its behavior. When the algorithm is on a given node, the simulation system only needs to keep track of two global variables: the node's current $n$-bit state and an array of size $h$ containing intermediate amplitude sums for each level of the tree. As it travels along various edges of the tree, the $n$-bit state is constantly rewritten according to these movements and state changes. Furthermore, the $h$-length array serves as a memory cache for previously computed intermediate amplitudes where the first array entry contains the entire circuit amplitude and successive entries contain the intermediate amplitudes of recursive steps. By periodically adding the ($n+1$)th entry to the $n$th entry after recursive calls finish, parent calls add up the amplitudes of child calls in lower layers until the first array entry contains the final amplitude. Combined with a few constant-size arguments passed within each recursive call, the total space usage is $O(n+h)$.
\par 
	Next, the time complexity of the algorithm can be proven as $O(t2^h)$ (where $t$ equals the number of non-branching gates). When the algorithm makes a transition along an edge of the tree, a classical manipulation of the bit register and the update of a phase factor occur in $O(1)$ time. Therefore, the algorithm takes time proportional to the number of edges in the tree, as each edge is traversed exactly twice: once advancing forwards through the circuit, and another time going ``backwards'' to reset the global $n$-bit register to its state prior to the edge for future operations. In a circuit of $t$ non-branching gates and $h$ branching gates, the worst-case (i.e.\ highest) edge count occurs when all branching gates happen first. In this case, the portion of the circuit containing the branching has $2^{h+1}$ edges, while the portion post-branching has $t2^h$ edges where the non-branching gates stretch forwards in straight lines extending from each of the $2^h$ nodes produced from the branching portion. Therefore, the sum of these worst-case edge counts gives a worst-case time performance of $O(2^{h+1} + t2^h) = O((t+2)2^h) = O(t2^h)$; an example of such a worst-case tree is shown in Figure 2. 
\begin{figure}[!b]
\centering
\includegraphics[width=0.7\textwidth]{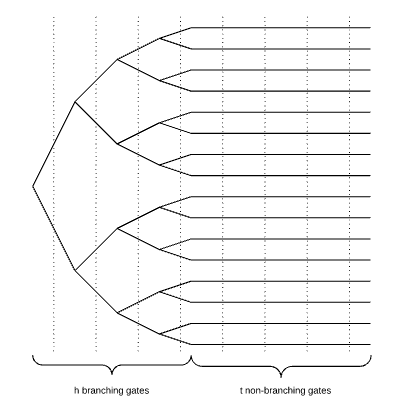}
\caption{the worst-case tree of maximum edge count.}
\end{figure}
\par
\noindent However, it should be noted that in most circuits the time complexity will not approach this upper bound, as the worst-case structure can be avoided by relegating the non-branching section to classical post-processing. Instead, the actual time taken will depend on the order of branching and non-branching gates within the circuit.
\par
	An effective optimization which can further improve performance in many cases lies in pruning the tree based on early termination of branches whose states cannot reach $endState$ using the remaining gates. At any node within the tree, if there are less gates left than the number of differing bits between the node's state and $endState$, then computation can be prematurely terminated; from that point onwards, reaching $endState$ is impossible due to each gate being capable of only changing---at most---one bit in the node state. Therefore, this pruning adjustment speeds up simulation by a varying factor with no sacrifice of accuracy, with the number of edges pruned depending on specific parameters of the simulated circuit. Figure 3 shows this optimization applied to the example introduced in Figure 1.
\begin{figure}[!b]
\centering
\includegraphics[width=0.48\textwidth]{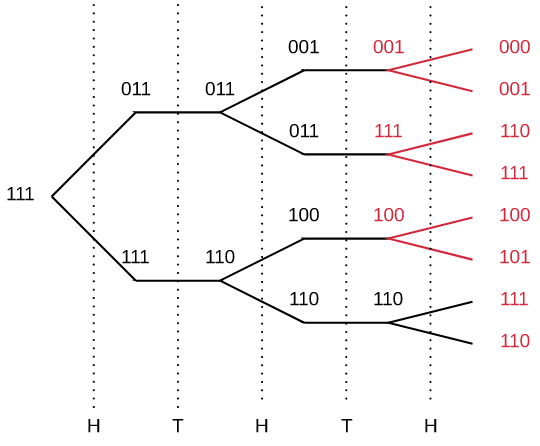}
\caption{an application of the pruning optimization on Figure 1's circuit in the calculation of \melem{010}{C}{111}, where nodes/edges outlined in red are no longer traversed.}
\end{figure}
\par
	To conclude the presentation of the algorithm's mechanism, the pseudocode below summarizes this section's discussed features and processes.
\begin{algorithm}[!]
\caption{Calculate \melem{endState}{C}{startState}}
\begin{algorithmic}
\STATE \textbf{Global Vars:} \textit{currState, ampRegister, endState, startState}
\newline
\STATE \textbf{recursiveStep(}\textit{depth}, \textit{phase}\textbf{):} //recursive method
\WHILE{next gate exists}
	\STATE{read next gate from \textit{C}}
	\IF{\textit{endState} unreachable}
		\STATE \textit{ampRegister}[\textit{depth}] := 0 //pruning adjustment
	\ELSE 
		\IF{non-branching}
			\STATE update \textit{phase}
			\STATE update \textit{currState}
		\ELSE 
			\STATE \textit{ampRegister}[\textit{depth}] := 0 //before any branches computed
			\FORALL{branches}
				\STATE \textit{currState} := branch state //move ``into'' branch's state
				\STATE \textbf{recursiveStep(}\textit{depth}+1, \textit{newPhase}\textbf{)} //recursive call for this branch
				\STATE \textit{ampRegister}[\textit{depth}] += \textit{ampRegister}[\textit{depth}+1] //add branch's amplitude to parent call's value
				\STATE \textit{currState} := original state //revert state for future computation
			\ENDFOR
		\ENDIF
	\ENDIF
\ENDWHILE
\IF{\textit{currState} = \textit{endState}}
	\STATE \textit{ampRegister}[\textit{depth}] := \textit{phase} //path reached endState
\ELSE
	\STATE \textit{ampRegister}[\textit{depth}] := 0
\ENDIF
\newline

\STATE \textbf{main():} //main simulation method
\STATE \textit{currState} := \textit{startState}
\STATE \textbf{recursiveStep(0, 1)} //top recursive call
\RETURN ampRegister[0] //return the top-level amplitude
\end{algorithmic}
\end{algorithm}

\section{Performance Analysis and Comparisons}
	To demonstrate the memory and time performances analyzed in the previous section, this section compares the recursive path-summing algorithm's performance to other general purpose simulation algorithms' using three types of benchmark circuits. Specifically, tests were conducted on the state vector simulation method and a circuit layer slicing approach developed by Aaronson and Chen \cite{aaronson} using $O(n(2d)^{n + 1})$ time and $O(n\log{d})$ space (where $d$ is a ``depth'' variable directly related to the circuit length) in addition to the paper's algorithm . A fourth simulation algorithm based on probabilistic circuit stepping named SEQCSim \cite{SEQCSim} was also noted; however, difficulty in understanding and executing its documented implementation resulted in the exclusion of its tests. Specifically, its online program files had some computing architecture-specific aspects which were difficult to adapt to our testing machine, and only a rudimentary understanding of the algorithm was achieved. However, this does not preclude the possibility of SEQCSim being a viable simulation method, which future reports could show.
\par
	The three circuit types chosen for performance benchmarks---layered Hadamard transform, layered quantum Fourier transform, and HSP (Hidden Subgroup Problem) standard method circuits, were selected as practical examples of quantum algorithms which will be commonly run on future quantum computers. Quantum algorithms known to achieve exponential speedups over their classical counterparts---including Shor's factoring algorithm \cite{shor} and Simon's algorithm \cite{simon}---often begin and end with Hadamard or quantum Fourier transforms, computing some classical function in between the two transforms. More generally, quantum algorithms which have attracted the most attention to date solve specific cases of a general problem called the hidden subgroup problem. The so-called "standard" method of solving the hidden subgroup problem \cite{hsp} is thus extremely relevant to the field of quantum computing; for this reason, a version of the method reminiscent of Shor's algorithm is simulated. 
\par
	Layered Hadamard transform circuits used for benchmarks consisted of two $n$-bit Hadamard transforms surrounding $n$ randomly generated Toffoli gates (universal for classical computation), resulting in $2n$ branching and $n$ non-branching gates. Additionally, layered quantum Fourier transform circuits contained $2n$ branching and $2(n(n-1)/2)+n$ non-branching gates, using a quantum Fourier transform circuit with $n(n+1)/2$ gates \cite{nielsen}. In the standard method test circuit used, the $n$-qubit register is separated into $a$ and $b$ registers of respective sizes $\lfloor2n/3\rfloor$ and $n-\lfloor2n/3\rfloor$; the register sizes were chosen based on the worst-case sizes of Shor's algorithm, where the integer input is of size $O(2^b)$ (and thus of length $b$) and the $a$ register's size varies from $b$ to $2b$ depending on the value of $b$. Using these registers, the test circuit consists of a Hadamard transform on register $a$, $n$ random Toffoli gates controlled by $a$ onto $b$, and a quantum Fourier transform on $a$, using $2\lfloor2n/3\rfloor$ branching and $\lfloor2n/3\rfloor(\lfloor2n/3\rfloor-1)/2 + n$ non-branching gates. However, executions for different register sizes (for example, when $\lvert a \rvert = \lvert b \rvert$) were not tested and remain an interesting open problem for researchers to simulate and collect data on.
\par
	Testing was done using a C++ implementation created as the PocketSimulator project \cite{code}.  All executions were performed on an ``Ivy Bridge'' 2.5 GHz Intel ``Core i5'' processor (3210M) with 8 GB of RAM; the test results are displayed in Figure 4, where each data point was calculated as an average of three trials.
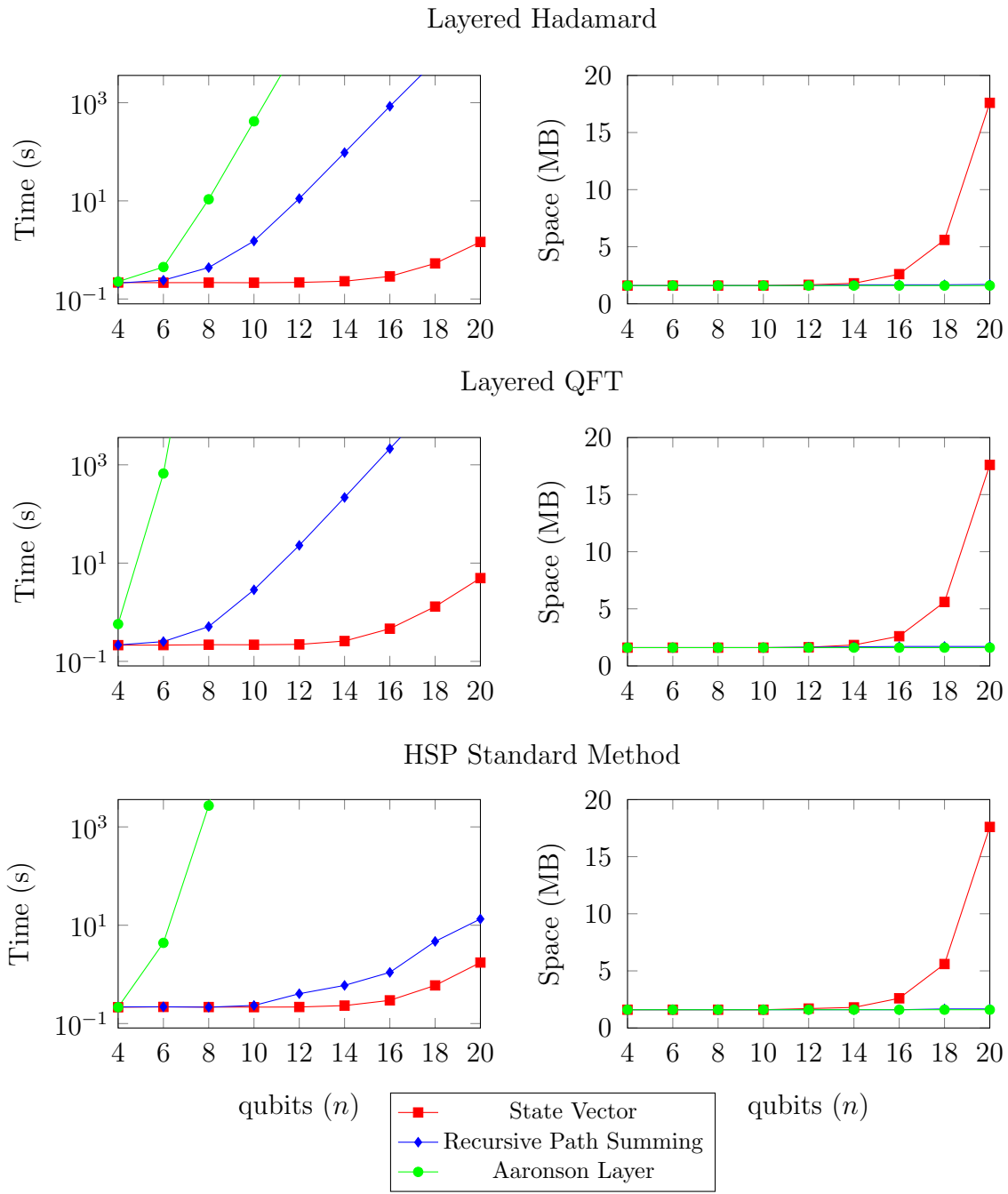
\begin{figure}[]
\centering
\begin{tikzpicture}
\begin{groupplot} [group style={group size=2 by 3, horizontal sep=2.2cm, vertical sep=2cm}, height=5cm, width=7cm, xtick = data
]
\nextgroupplot[
ymode=log,
title=Layered Hadamard,
title style={xshift = 20ex, yshift = 1.5ex}, 
ylabel={Time (s)},
ylabel style={yshift=-.5ex},
xmin = 4, xmax = 20,
x tick label style = {yshift = -0.5ex},
ymax = 3600,
]
\addplot [color=red, mark=square*] table {h_time_vec.dat};
\addplot [color=blue, mark=diamond*] table {h_time_pocket.dat};
\addplot [color=green, mark=otimes*] table {h_time_aaron.dat};

\nextgroupplot[
ylabel={Space (MB) },
xmin = 4, xmax = 20,
x tick label style = {yshift = -0.5ex},
ymin = 0, ymax = 20,
y tick label style = {xshift = -0.5ex},
]
\addplot [color=red, mark=square*] table {h_space_vec.dat};
\addplot [color=blue, mark=diamond*] table {h_space_pocket.dat};
\addplot [color=green, mark=otimes*] table {h_space_aaron.dat};

\nextgroupplot[
ymode = log,
title= Layered QFT,
title style={xshift=20ex, yshift=1.5ex},
xmin = 4, xmax = 20,
x tick label style = {yshift = -0.5ex},
ylabel={Time (s)}, ylabel style={yshift=-.5ex},
ymax = 3600,
]
\addplot [color=red, mark=square*] table {lqft_time_vec.dat}; 
\addplot [color=blue, mark=diamond*] table {lqft_time_pocket.dat}; 
\addplot [color=green, mark=otimes*] table {lqft_time_aaron.dat};

\nextgroupplot[
ylabel={Space (MB)},
xmin = 4, xmax = 20,
x tick label style = {yshift = -0.5ex},
ymin = 0, ymax = 20,
y tick label style = {xshift = -0.5ex},
]
\addplot [color=red, mark=square*] table {lqft_space_vec.dat};
\addplot [color=blue, mark=diamond*] table {lqft_space_pocket.dat};
\addplot [color=green, mark=otimes*] table {lqft_space_aaron.dat};

\nextgroupplot[
ymode=log,
title=HSP Standard Method,
title style={xshift=20ex, yshift=1ex}, 
xlabel={qubits ($n$)},
xlabel style={yshift=-1ex},
ylabel={Time (s)},
xmin = 4, xmax = 20,
x tick label style = {yshift = -0.5ex},
ymax = 3600,
legend style={at={(1.2,-0.5)}, anchor=center}
]
\addplot [color=red, mark=square*] table {hsp_time_vec.dat};
\addplot [color=blue, mark=diamond*] table {hsp_time_pocket.dat};
\addplot [color=green, mark=otimes*] table {hsp_time_aaron.dat};
\addlegendentry{State Vector};    
\addlegendentry{Recursive Path Summing};    
\addlegendentry{Aaronson Layer};

\nextgroupplot[
title style={yshift=1ex}, 
xlabel={qubits ($n$)}, 
xlabel style={yshift=-1ex},
ylabel={Space (MB)},
xmin = 4, xmax = 20,
x tick label style = {yshift = -0.5ex},
y tick label style = {xshift = -0.5ex},
ymin = 0, ymax = 20,
]
\addplot [color=red, mark=square*] table {hsp_space_vec.dat};
\addplot [color=blue, mark=diamond*] table {hsp_space_pocket.dat};
\addplot [color=green, mark=otimes*] table {hsp_space_aaron.dat};

\end{groupplot}
\end{tikzpicture}
\caption{memory and time graphs for all tested circuit types, tested on qubit values ranging from 4 to 20; time graphs were capped at a maximum of 1 hour (3600 seconds).}
\end{figure}

\vspace{16mm}
\par
	From Figure 4, a drastic difference in space usage for all three circuits can be seen between the state vector approach's exponential curve and the other two algorithms' lines, confirming the theoretical space complexities explained earlier. In the layered Hadamard circuit tests, the recursive path-summing algorithm beats the layer-slicing approach by orders of magnitude in time due to the growing exponent base $d$ in the latter's time complexity, but still grows much faster than the state vector simulation; in the layered quantum Fourier transform tests, all three algorithms take even more time due to the extra gates present in the Fourier transform.
\par
	However, for the HSP standard method, the path-summing algorithm is almost able to match up to the state vector approach in running time as $n$ grows larger! These results can be explained using the conjectured time complexities of each algorithm and each circuit's properties: for the layered circuits,
$$ h = 2n \longrightarrow t2^h \gg (t+h)2^n $$
\noindent which accounts for the state vector simulation's faster performance in these cases. On the other hand, for the HSP standard method, 
$$ h = 2\lfloor2n/3\rfloor \longrightarrow t2^h > (t+h)2^n $$
\noindent which explains the recursive path-summing algorithm's relative speed increase, which may have been even greater if $\lvert a \rvert < 2\lvert b \rvert$ where less branching gates are used. Additionally, looking at the algorithms' time complexities shows that even if $t > O(n)$, the runtimes stay relatively similar; thus, the $n$ Toffoli gates used in our testing implementation could be changed to compute any classically controlled function with respect to registers $a$ and $b$ while preserving the runtime relationship between the path integral and state vector algorithms.
\par
	Despite the gaps in speed between the algorithms for various circuits, all time graphs indicate exponential growth as conjectured. However, in general, the performance tests conducted reveal that the recursive path-summing implementation is able to provide satisfactory time performance for the hidden subgroup problem standard method (because $n$ is close to $h$) while using virtually no space by comparison, a major result of this paper.

\section{Conclusion}
	Experimentally, the presented path integral implementation uses linear space and simulates certain circuits faster than existing low-space methods; for certain circuits, it uses time comparable to the state vector approach while using almost no space relative to exponential space requirements. However, execution tests may require further refinement to reach a robust conclusion in this regard. In particular, implementations for both the layered-circuit and state vector methods were written specifically to benchmark the paper's implementation and may have contained bugs resulting in slower performance; memory and runtime data may have also been inaccurately collected due to machine or collection program flaws. We invite other researchers to peruse the implementations used in the paper \cite{code} and investigate these possibilities.
\par
	Most interestingly, the results of this paper show that the recursive path-summing algorithm is able to sacrifice some runtime at the great benefit of using linear space in contrast to exponential space required by the state vector approach, for certain circuits such as the HSP standard method. This result implies that when testing supremacy on the family of HSP standard method circuits, the recursive path-summing algorithm is a more accurate method than the state vector approach by which to determine the supremacy threshold on larger numbers of qubits. From the experiments presented, it is not immediately clear whether the algorithm is able to simulate greater than 50 qubits on the HSP; however, considering the moderate $\approx20$ qubits tests where runtime was close to the state vector's, it is definitely a possibility.
\par	
	There are several theoretical open problems regarding the algorithm's conceptual basis and the general subfield of classical simulation which could be investigated with future research. For one, the structure of the algorithm and its performance on the HSP standard method indicates the possibility of obtaining a more efficient algorithm. Specifically, the question of whether the runtime complexity of the algorithm can be improved from $O(t2^h)$ to $O(t2^d)$---where $d$ is the circuit depth of the simulated algorithm---may lead to a faster space-efficient algorithm or better understanding of the algorithm's theoretical implications. More generally, the open problem of proving lower bounds on the time and memory requirements for classical simulation is also closely related to the concept of quantum supremacy and the inherent power of quantum computation. Finding such a definitive lower bound would be an even greater step in determining where the boundary of quantum supremacy lies. 
\vspace{8mm}
\par
	The algorithm implementation presented in this paper is an alternative simulation method which eliminates the need for large memory registers such as those required by the state vector method. From the presented results, the recursive path-summing algorithm is able to establish itself as a viable candidate for general simulation purposes. More specifically, the HSP tests also show that current state vector memory-based limitations of the supremacy frontier may not hold for the HSP family. As the field of quantum computing continues to expand and physical quantum computers approach this threshold, this paper and its simulation implementation is another step towards fully understanding the power of quantum computing. 

\section*{Acknowledgments}
I would like to thank my uncle, Yaoyun Shi, for his invaluable advice and suggestions as well as my dad, Yanyun Shi, for his encouragement and guiding perspective.

\bibliographystyle{abbrv}
\raggedright
\bibliography{sources}

\begin{thebibliography}{10}

\bibitem{aaronson}
S.~{Aaronson} and L.~{Chen}.
\newblock {Complexity-Theoretic Foundations of Quantum Supremacy Experiments}.
\newblock {\em ArXiv e-prints}, Dec. 2016.

\bibitem{newstabilizer}
S.~{Aaronson} and D.~{Gottesman}.
\newblock {Improved simulation of stabilizer circuits}.
\newblock {\em Physical Review A}, 70(5):052328, Nov. 2004.

\bibitem{feynman}
R.~Feynman and P.~W. Shor.
\newblock Simulating physics with computers.
\newblock {\em SIAM Journal on Computing}, 26:1484--1509, 1982.

\bibitem{SEQCSim}
M.~P. {Frank}, U.~H. {Meyer-Baese}, I.~{Chiorescu}, L.~{Oniciuc}, and R.~A.
  {van Engelen}.
\newblock {Space-Efficient Simulation of Quantum Computers}.
\newblock {\em ArXiv e-prints}, Dec. 2008.

\bibitem{stabilizer}
D.~{Gottesman}.
\newblock {The Heisenberg Representation of Quantum Computers}.
\newblock {\em eprint arXiv:quant-ph/9807006}, July 1998.

\bibitem{largesim}
T.~{H{\"a}ner}, D.~S. {Steiger}, M.~{Smelyanskiy}, and M.~{Troyer}.
\newblock {High Performance Emulation of Quantum Circuits}.
\newblock {\em ArXiv e-prints}, Apr. 2016.

\bibitem{smallquantcomp}
J.~A. Jones and M.~Mosca.
\newblock Implementation of a quantum algorithm on a nuclear magnetic resonance
  quantum computer.
\newblock {\em The Journal of Chemical Physics}, 109(5):1648--1653, 1998.

\bibitem{hsp}
C.~{Lomont}.
\newblock {The Hidden Subgroup Problem - Review and Open Problems}.
\newblock {\em eprint arXiv:quant-ph/0411037}, Nov. 2004.

\bibitem{tensor}
I.~L. Markov and Y.~Shi.
\newblock Simulating quantum computation by contracting tensor networks.
\newblock {\em SIAM J. Comput.}, 38(3):963--981, June 2008.

\bibitem{nielsen}
M.~A. Nielsen and I.~L. Chuang.
\newblock {\em Quantum Computation and Quantum Information: 10th Anniversary
  Edition}.
\newblock Cambridge University Press, New York, NY, USA, 10th edition, 2011.

\bibitem{supremacy}
J.~{Preskill}.
\newblock {Quantum computing and the entanglement frontier}.
\newblock {\em ArXiv e-prints}, Mar. 2012.

\bibitem{code}
A.~Shi.
\newblock {PocketSimulator}.
\newblock \url{https://github.com/AShiTheCoder/PocketSimulator}, 2017.

\bibitem{shor}
P.~W. Shor.
\newblock Polynomial-time algorithms for prime factorization and discrete
  logarithms on a quantum computer.
\newblock {\em SIAM J. Comput.}, 26(5):1484--1509, Oct. 1997.

\bibitem{simon}
D.~R. Simon.
\newblock On the power of quantum computation.
\newblock {\em SIAM J. Comput.}, 26(5):1474--1483, 1997.

\bibitem{pfaffian}
L.~G. Valiant.
\newblock Quantum computers that can be simulated classically in polynomial
  time.
\newblock In {\em Proceedings of the Thirty-third Annual ACM Symposium on
  Theory of Computing}, STOC '01, pages 114--123, New York, NY, USA, 2001. ACM.

\bibitem{quidd}
G.~F. {Viamontes}, I.~L. {Markov}, and J.~P. {Hayes}.
\newblock {Improving Gate-Level Simulation of Quantum Circuits}.
\newblock {\em eprint arXiv:quant-ph/0309060}, Sept. 2003.

\bibitem{limitedentangle}
G.~{Vidal}.
\newblock {Efficient Classical Simulation of Slightly Entangled Quantum
  Computations}.
\newblock {\em Physical Review Letters}, 91(14):147902, Oct. 2003.

\end{thebibliography}

\end{document}